\documentclass[usegraphicx]{mn2e}

\newcommand       \be           {\begin{equation}}
\newcommand       \ee           {\end{equation}}

\begin{document}

\title[Electric dipole moments and disalignment of grains]
{Electric dipole moments and disalignment of interstellar dust grains}
\author[M. E. Jordan \& J. C. Weingartner]
{Margaret E. Jordan\thanks{Email: mjordan4@gmu.edu; jweinga1@gmu.edu} \& 
Joseph C. Weingartner\footnotemark[1]\\
Department of Physics and Astronomy, George Mason University,
MSN 3F3, 4400 University Drive, Fairfax, VA 22030, USA}

\maketitle

\begin{abstract}

The degree to which interstellar grains align with respect to the interstellar
magnetic field depends on disaligning as well as aligning mechanisms.  
For decades,
it was assumed that disalignment was due primarily to the random angular 
impulses a grain receives when colliding with gas-phase atoms.  Recently, 
a new disalignment mechanism has been considered, which may be very potent
for a grain that has a time-varying electric dipole moment and drifts across
the magnetic field.  We provide quantitative estimates of the disalignment
times for silicate grains with size $\ga 0.1 \micron$.  These appear to
be shorter than the time-scale for alignment by radiative torques, unless
the grains contain superparamagnetic inclusions.  

\end{abstract}


\section{INTRODUCTION}

Observations of starlight polarization have revealed that some 
interstellar dust
grains are non-spherical and aligned.  The degree of alignment, and hence the 
polarization, depends on both aligning processes (e.g., radiative torques
and paramagnetic dissipation)
and disaligning processes (e.g., random torques arising from collisions with 
gas atoms).  See Whittet (2004) for a review of polarization observations and 
Lazarian (2003), Roberge (2004), and Lazarian (2007) for reviews of alignment 
theory.  

Recently, Weingartner (2006, hereafter W06) proposed an alternative 
disalignment mechanism
for a grain that has a time-varying electric dipole moment $\bmath{p}$ and 
drifts across the interstellar magnetic field.  The potency of this mechanism
is sensitive to the magnitude and time-scale of fluctuations in $\bmath{p}$. 
W06 considered highly simplified models for the fluctuating
electric dipole moment.  Here, we examine this process and the implications
for disalignment in greater detail.  We will consider relatively large
(size  $\ga 0.1 \ \micron$) silicate grains, since the $9.7$ and 
$20 \micron$ features exhibit polarization (e.g., Martin \& Whittet 1990; Smith
et al. 2000; Wright et al. 2002) and the wavelength dependence of the 
observed polarization implies that relatively small grains are not efficient
polarizers (Kim \& Martin 1995).

In \S \ref{sec:summary}, we review the main elements of disalignment 
associated with fluctuations in $\bmath{p}$.  Next, we introduce models
for the transport of charge to and within silicate grains 
(\S \ref{sec:models}).  We describe simulations of the fluctuating dipole
moment and associated disalignment in \S \ref{sec:simulations} and present
results in \S \ref{sec:results} and conclusions in \S \ref{sec:conclusion}.

\section{DISALIGNMENT ASSOCIATED WITH TIME-VARYING 
ELECTRIC DIPOLE MOMENTS}
\label{sec:summary}

When a gas atom collides with a grain, it imparts an angular impulse to
the grain.  If no other mechanisms excite rotation, then the energy in 
rotation about any axis is $\sim \frac{1}{2} k_{\mathrm{B}} T_{\mathrm{gas}}$, 
where $k_{\mathrm{B}}$ is Boltzmann's constant and $T_{\mathrm{gas}}$ is the 
gas temperature. 
Such motion is called `thermal rotation'.  The thermal rotation rate for
a sphere with radius $a$ is given by
\begin{eqnarray}
\label{eq:omega_T}
\nonumber
\omega_{T} & = & \left( \frac{15 k_{\mathrm{B}} T_{\mathrm{gas}}}
{8 \pi \rho a^5} \right)^{1/2}\\
\nonumber
& = &
1.66 \times 10^5 \, \left( \frac{\rho}{3 \ \mathrm{g} \ \mathrm{cm}^{-3}} 
\right)^{-1/2}
\left( \frac{T_{\mathrm{gas}}}{100 \ \mathrm{K}} \right)^{1/2}\\ 
& & \times \left( \frac{a}{0.1 \ \micron}
\right)^{-5/2} \ \mathrm{s}^{-1} ,
\end{eqnarray}
where $\rho$ is the density of the grain material.
In general, grains are subjected to additional torques that may drive them
to suprathermal rotation, with angular speed $\omega > \omega_{T}$ 
(Purcell 1975, 1979; Draine \& Lazarian 1998).  
For thermally rotating grains, the random collisional
impulses constitute an important disalignment mechanism.

A spinning grain with non-zero electric charge acquires a magnetic dipole
moment $\bmu \parallel \bomega$ (Martin 1971).  
Dolginov \& Mytrophanov (1976) showed that the Barnett effect
(i.e., the tendency for a spinning paramagnetic solid to
acquire a magnetization parallel or anti-parallel to 
$\bomega$) can provide a much larger moment.  Specifically,
the Barnett magnetic moment $\bmu_{\rm Bar} = \chi_0 \bomega V/\gamma_g$, where
$\chi_0$ is the static magnetic susceptibility, $\gamma_g$ is the
gyromagnetic ratio of the microscopic magnetic dipoles that are responsible
for the grain's paramagnetism, and $V$ is the grain volume.
The susceptibility depends on the number density of paramagnetic ions or 
nuclei in the grain material and is thus rather uncertain.  
We will adopt $\chi_0 \sim 5 \times 10^{-3} (T_d / 15 \ \mathrm{K})^{-1}$, 
where $T_d$ is the grain temperature (Draine 1996; W06).  With this estimate, 
the Barnett magnetic moment for silicate grains is
\begin{eqnarray}
\label{eq:mu_Bar_sil}
\nonumber
|\mu_{\mathrm{Bar}}|(\mathrm{sil}) & \approx &
1.2 \times 10^{-19} \, \left( \frac{T_d}{15 \ \mathrm{K}}
\right)^{-1} \left( \frac{a}{0.1 \ \micron} \right)^3\\
& & \times \left( \frac{\omega}
{10^5 \ \mathrm{s}^{-1}} \right) \ 
\mathrm{statC} \ \mathrm{cm}.
\end{eqnarray}

Suppose $\bmu$ is parallel or anti-parallel to the grain's angular momentum
vector $\bmath{J}$:  $\bmu = \mu_J \bmath{\hat{J}}$.
The magnetic torque $\bmath{{\Gamma}_{\mu}} = \bmu \bmath{\times} \bmath{B}$ 
causes
$\bmath{J}$ to precess about the interstellar magnetic field $\bmath{B}$ at 
rate
\begin{eqnarray}
\label{eq:prec}
\nonumber
|\Omega_0| & = & \frac{|\mu| B}{J}\\
& \approx & 314 \,  \left(
\frac{|\mu|}{10^{-19} \ \mathrm{statC} \ \mathrm{cm}} \right) 
\left( \frac{\rho}{3 \ \mathrm{g} \ 
\mathrm{cm}^{-3}} \right)^{-1}\\
\nonumber
& & \times \left( \frac{B}{5 \  \mu \mathrm{G}} \right)
\left( \frac{a}{0.1 \ \micron} \right)^{-5} \left( \frac{\omega}
{10^5 \ \mathrm{s}^{-1}} \right)^{-1} \ \mathrm{yr}^{-1}.
\end{eqnarray}
Since $\mu \propto \omega$, $\Omega_0$ is independent of $\omega$.
The combination of an aligning torque (e.g., the radiative torque) and 
the magnetic torque drives the grain towards
rapid precession with a constant precession angle
$\theta_{\rm align}$ (i.e., $\theta_{\rm align}$ is the angle between 
$\bmath{B}$ and $\bmath{J}$).  A large ensemble of grains will be 
characterized by a uniform distribution in precession phase.  As a result,
the observed starlight polarization is either parallel or perpendicular to 
$\bmath{B}$.  (If $\theta_{\rm align} = 0$ and $\bmath{J} \parallel 
\bmath{\hat{a}_1}$, the grain principal axis of greatest moment of inertia, 
then the polarization $\parallel \bmath{B}$.)  Note that 
alignment of the grain body with respect to $\bmath{J}$ is also a
necessary condition for polarization.  

If a grain has an electric dipole moment $\bmath{p}$ and drifts with 
velocity $\bmath{v}$ across
$\bmath{B}$, then it experiences a torque 
$\bmath{\Gamma_p} = \bmath{p} \bmath{\times} (\bmath{v} \bmath{\times} 
\bmath{B})/c$
in addition to the magnetic torque ($c$ is the speed of light).  
If $\bmath{p} = p_J \bmath{\hat{J}}$,
then the grain precesses about an axis tilted at angle 
$\delta = \tan^{-1} |\Upsilon|$ relative to $\bmath{B}$ and the precession
rate is increased by the factor $(1+\Upsilon^2)^{1/2}$, where
\be
\label{eq:Upsilon}
\Upsilon \equiv \frac{p_J v_{\perp}}{\mu_J c}
\ee
with $v_{\perp}$ the component of $\bmath{v}$ transverse to $\bmath{B}$
(W06).  

If $\bmu$ and $\bmath{p}$ are not parallel or anti-parallel to $\bmath{J}$,
then the magnetic and electric torques must be averaged over the 
extremely rapid grain rotation.  The resulting dynamics is identical to 
that for which $\bmath{J}$, $\bmu$, and $\bmath{p}$ all lie along 
$\bmath{\hat{a}_1}$, except with the following substitutions in equation 
(\ref{eq:Upsilon}):
\be
\label{eq:mu_J}
\mu_J = \left(\frac{\mu}{\omega} \right) \frac{qJ}{I_1}~~~,
\ee
\be
\label{eq:p_J}
p_J = \pm \left( \bmath{p} \bmath{\cdot} \bmath{\hat{a}_i} \right) f_i(q)~~~,
\ee
with $I_i$ the moment of inertia along $\bmath{\hat{a}_i}$ (W06).  
Both $\mu_J$ and $p_J$ depend on the grain's rotational state through the
parameter $q \equiv 2 I_1 E/J^2$ ($E$ is the rotational energy).  
In equation 
(\ref{eq:p_J}), the + (-) sign is selected when 
$\bmath{J} \bmath{\cdot} \bmath{\hat{a}_i}$ $> 0$ ($<0$) and the factor 
$f_i(q)$
is given in eq. 9 of W06.  (The choice of $i$ is also discussed following
eq. 9 in W06.)  
If the Barnett effect is responsible for the magnetic dipole moment,
then $(\mu/\omega) = \chi_0 V/\gamma_g$.

If $\Upsilon$ is constant in time, then the only consequence of the electric 
dipole is to tilt the precession axis relative to the magnetic field direction.
However, $\Upsilon$ can vary on relatively short time-scales.  
W06 discussed two
sources of variation:  1.  Upon each discrete charging event (e.g., the 
capture of an electron from the gas or photoejection of an electron), 
$p_J$ changes.  2.  When the grain's rotational state ($q$ and/or flip state)
varies, $p_J/\mu_J$ varies (eqs. \ref{eq:mu_J} and \ref{eq:p_J}).  

Two processes can yield rapid variations in the grain rotational state:
1.  Thermal fluctuations, in which energy is exchanged between grain rotation
and vibrational modes (Lazarian 1994; Lazarian \& Roberge 1997; Lazarian \& 
Draine 1997, 1999a, 1999b; Weingartner 2009).
2.  Collisions with gas-phase atoms, which can stick to, reflect from, or 
evaporate from the surface, perhaps after forming a molecule (Hoang \& 
Lazarian 2009).  The efficacy 
of both of these mechanisms drops off dramatically as the grain rotation 
becomes suprathermal.  

In this paper, we will only consider variations in $p_J$ associated with 
discrete charging events.  We also will assume $q=1$ in equation 
(\ref{eq:mu_J}) and $f_i(q)=1$ in equation (\ref{eq:p_J}), which are good
approximations for suprathermally rotating grains.    
Thus, the analysis presented here is not complete for thermally rotating
grains.

Since the charging processes are stochastic processes, $\Upsilon$ varies 
stochastically, yielding random variations in the precession axis.  Each 
time the precession axis changes direction, the precession angle changes.
When these events occur at random precession phases, 
$\theta_{\rm align}$ varies stochastically.  In other words, the 
grain experiences disalignment.  

W06 considered a simple scenario in which $\Upsilon$ has constant magnitude
but stochastically reverses sign, on time-scale $\tau_{\rm flip}$, finding
the following approximations for the disalignment time-scale when 
$\tau_{\rm flip}$ is short or long compared with the precession time-scale:
\be
\label{eq:tau_dis1}
\tau_{\rm dis} \sim \Upsilon^{-2} |\Omega_0|^{-2} 
\tau_{\rm flip}^{-1}~,~~{\rm if}~~~\tau_{\rm flip}
\ll |\Omega_0|^{-1} (1+\Upsilon^2)^{-1/2}
\ee
\be
\label{eq:tau_dis2}
\tau_{\rm dis} \sim (1+\Upsilon^{-2})
\tau_{\rm flip}~,~~{\rm if}~~~\tau_{\rm flip} \gg
|\Omega_0|^{-1} (1+\Upsilon^2)^{-1/2}.
\ee
In the following section, we will consider more detailed models for the 
fluctuating electric dipole moment.  

\section{GRAIN CHARGING MODELS}
\label{sec:models}

A grain charging model that follows the evolution of the electric dipole
moment $\bmath{p}$ must treat both the processes that deliver charge to 
the grain and those that transport charge within the grain.  
In the cold, neutral, interstellar medium, the dominant charge delivery  
mechanisms are starlight-induced photoelectric emission and sticking 
collisions of gas-phase electrons.  

\subsection{Idealizations for Charge Transport Within a Grain}
\label{sec:idealizations}

Bulk, neutral silicates are good insulators, with a full valence band and 
empty conduction band.  Observations of the $9.7 \ \micron$ band profile
indicate that interstellar silicates are predominantly amorphous 
(Li \& Draine 2001; Kemper, Vriend, \& Tielens 2004; Li, Zhao, \& Li 2007).
In amorphous materials, localized energy states (`traps')
appear in the tails of the conduction and valence bands.
For any realistic interstellar grain, there are also localized states
associated with impurity atoms.  Electrons and holes can hop from site to
site with assistance from a phonon (e.g., Mott \& Davis 1971; Blaise 2001),
so no grain is perfectly insulating.

The rate at which an electron hops from site $i$ to site $j$ is 
typically approximated as
\be
\label{eq:R_hop}
R_{\mathrm{hop}} = \nu_{\mathrm{ph}} \exp(-2 r_{ij} / d_{\psi})
\exp(-W_{i \rightarrow j}/k_{\mathrm{B}} T_{\mathrm{d}})
\ee
(Ambegaokar et al.~1971; Mady et al.~2007), 
where $\nu_{\mathrm{ph}} \sim 10^{13} \ \mathrm{s}^{-1}$ is the phonon 
frequency (Brucato et al.~2002),
$r_{ij}$ is the distance between sites $i$ and $j$,
$d_{\psi}$ is the electron localization length,
$W_{i \rightarrow j} = \max[E_j - E_i, 0]$,
$E_i$ is the electron energy when localized at site $i$,  
and $T_{\mathrm{d}}$ is the dust temperature.

A completely rigorous treatment of the grain electric dipole moment would
include following the charges as they hop among traps.  However, this approach
is not feasible.  First, the quantities appearing in equation (\ref{eq:R_hop}),
namely $d_{\psi}$ and the trap energy distribution, are poorly known.  
Second, even for tight binding at traps (e.g., 
$d_{\psi} \approx 2 \ \mathrm{\AA}$),
there are typically numerous neighboring traps for which the hopping time
is orders of magnitude smaller than the time between discrete charging 
events (which itself is orders of magnitude smaller than the disalignment
time).  This is true even when a charge is well-localized within the vicinity
of a particularly deep trap.  Given the large disparity in time-scales, the 
CPU time for a simulation that follows hopping in detail is prohibitive.

Fortunately, a few simple, plausible idealizations are available and do not
strain computational resources.  We will consider the following 4 models:

\noindent {\bf 1.  A perfect insulator.}  
Each time a charge arrives at the grain (either an
electron from the gas or a hole left following photoemission), it remains
at its arrival point forever.  The full hopping model simplifies to this 
case when $d_{\psi} \rightarrow 0$, if the typical distance between traps is 
much less than the grain size.  This idealization is also reasonable if 
(a) the typical distance between {\it deep} traps is much less than the grain 
size and (b) the deep traps effectively retain charges in their immediate 
vicinity.  That is, a charge is unlikely to leave the `sphere of influence' 
of a deep trap before recombining.

\noindent {\bf 2.  A perfect conductor.}  
The excess charge on the grain is completely 
delocalized.  For a homogeneous, spherical grain, the electric dipole moment
$\bmath{p}$ vanishes in this case.  For non-spherical shapes, $p \propto Z$, 
the net charge on the grain (in units of the proton charge).  
This model is probably not suitable for 
interstellar grains, since we expect $Z$ to be less than the total number
of deep traps in the grain.  Still, it is useful to consider this case, to
constrain the range of possible outcomes.

\noindent {\bf 3.  A conducting grain with deep traps.}
Some number of deep traps are located at random positions within the grain.
When a charge arrives, it immediately moves to the nearest available trap 
(either occupying it or recombining with a resident charge of the opposite 
sign).  This model approaches case (1) as the number density of deep traps
increases.  

\noindent {\bf 4.  A partially conducting grain with deep traps.}  
Same as (3), except that a charge executes a random walk through the grain, 
with 
some typical step size and frequency, until it comes close to an available
deep trap, where it gets stuck.  We assume that any adsorbates present on the 
grain surface are sufficiently dilute that there is no associated enhancement 
in conductivity along the surface.

\subsection{Collisional Charging}

The trajectories of charged particles in the vicinity of a grain with 
non-vanishing electric dipole moment $\bmath{p}$ differ from those for the
$\bmath{p} = 0$ case.  The distribution of arrival sites on the grain surface
is such as to reduce $p = |\bmath{p}|$.  Except for model (2) in 
\S \ref{sec:idealizations}, this effect is critical for limiting $p$.  
However, it is extremely difficult to treat for non-spherical grain shapes.
Thus, we will always treat the grain as a sphere when computing collisional
charging rates and the arrival sites of colliding particles.  For further 
simplification in these calculations, we also neglect the motion of the 
grain with respect to the gas.  Even though the grain's speed is assumed to be
roughly the
sound speed of the gas, the speed of the light electrons is greater by a 
factor $\approx (m_p/m_e)^{1/2}$ ($m_p$ and $m_e$ are the proton and electron
mass, respectively).  Thus, we do not expect this assumption to introduce 
serious error for electron collisional charging.  In addition, we neglect ion 
collisional charging, which is dominated by photoelectric emission.  These
simplifications are justified in Appendix \ref{app:drift}.

For a grain at rest with respect to the gas, the collisional charging rate is 
given by
\be
\label{eq:R_coll}
R = \pi a^2 n \, s \left( \frac{8 k_{\mathrm{B}} T_{\mathrm{gas}}}{\pi m} 
\right)^{1/2} \tilde{R}
\ee
where $n$ is the number density of the colliding
particles, $s$ is the sticking coefficient (i.e., the probability that the
particle sticks to the grain following a collision), 
$m$ is the mass of colliding particle, and $\tilde{R}$ accounts
for deviations of the collision cross section from the geometric cross 
section.  For the relatively large grains under consideration here, we 
adopt $s \approx 1/2$ (Weingartner \& Draine 2001, hereafter WD01).  
Draine \& Sutin (1987) provided expressions for $\tilde{R}$ for a charged,
conducting sphere, including the polarization of the grain by the charged
gas-phase particle.  The effect of polarization decreases with grain size
(as long as $T_{\mathrm{gas}}$ does not approach zero), and can be reasonably
neglected when $a \ge 0.1 \ \micron$.  

Consider a spherical grain with radius $a$ centered at the origin.  
Approximate the charge distribution within the grain as a point charge
$Q$ and point dipole $p \bmath{\hat{z}}$ ($p>0$) located at the origin.  
In spherical 
coordinates $(r, \theta, \phi)$, the electric force on a point charge $q$ is
\be
\bmath{F} = \frac{Qq}{r^2} \, \bmath{\hat{r}} + \frac{qp}{r^3} \left( 2 \cos 
\theta \, \bmath{\hat{r}} + \sin \theta \, \bmath{\hat{\theta}} \right)
\ee
and the potential is
\be
U = \frac{Qq}{r} + \frac{qp \cos \theta}{r^2}.
\ee
The equations of motion are
\be
\label{eq:r_dot_dot}
m \ddot{r} = m r \dot{\theta}^2  + m r \sin^2 \theta \, \dot{\phi}^2 + 
\frac{Qq}{r^2} + \frac{2qp \cos \theta}{r^3}
\ee
\be
m r \ddot{\theta} = -2 m \dot{r} \dot{\theta} + m r \sin \theta \cos \theta
\, \dot{\phi}^2 + \frac{qp \sin \theta}{r^3}
\ee
\be
\label{eq:phi_dot_dot}
m r \sin \theta \, \ddot{\phi} = -2 m \dot{r} \sin \theta \, \dot{\phi} - 
2 m r \cos \theta \, \dot{\theta} \dot{\phi}
\ee
where dots denote differentiation with respect to time and $m$ is the mass
of the point charge $q$.  

Employing Hamilton-Jacobi theory, we find the following conserved quantities:
\be
\label{eq:p_phi}
p_{\phi} \equiv m r^2 \sin^2 \theta \, \dot{\phi}
\ee
\be
\label{eq:beta}
\beta \equiv m^2 r^4 \dot{\theta}^2 + 2 m q p \cos \theta + 
\frac{p_{\phi}^2}{\sin^2 \theta}
\ee
\be
\label{eq:E}
E \equiv \frac{1}{2} m \dot{r}^2 + \frac{Qq}{r} + \frac{\beta}{2 m r^2} ,
\ee
as can be verified by direct time differentiation, substituting for the
second derivatives from equations 
(\ref{eq:r_dot_dot})--(\ref{eq:phi_dot_dot}).  

\begin{figure}
\includegraphics[width=84mm]{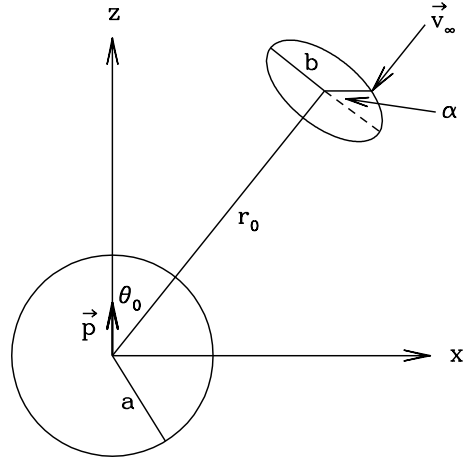}
\caption{
Parameters describing the trajectories of incoming charged particles.
        }
\label{fig:geometry}
\end{figure}

Our goals are to find (1) the rate at which incoming charged particles strike 
the grain surface and (2) the distribution of their arrival angles $\theta$,
given $Q$, $p$ and
$T_{\mathrm{gas}}$.  First, we describe the trajectory of
the incoming particle when it is still far from the grain (see Fig. 
\ref{fig:geometry}).   
Suppose its velocity is
\be
\label{eq:v_inf}
\bmath{v_{\infty}} = - v ( \cos \theta_0 \, \bmath{\hat{z}} + 
\sin \theta_0 \, \bmath{\hat{x}}).
\ee
The trajectory is offset from the line $x = \tan \theta_0 \, z$, which passes
through the grain center, by impact parameter $b$; angle $\alpha$ specifies
the displacement of the trajectory from the $x-z$ plane.  
Consider a plane front of incoming particles.  When the particle whose 
trajectory passes through the origin is located at distance $r_0$ from 
the origin, the coordinates of the other particles are
\be
(r, \theta, \phi) \approx \left( r_0, \theta_0 + \cos \alpha \frac{b}{r_0},
\frac{\sin \alpha}{\sin \theta_0} \frac{b}{r_0} \right)
\ee
and the components of their velocities are
\be
\label{eq:r_dot}
\dot{r} \approx - v
\ee
\be
\label{eq:theta_dot}
r \dot{\theta} \approx v \cos \alpha \, (b / r_0)
\ee
\be
\label{eq:phi_dot}
r \sin \theta \, \dot{\phi} \approx v \sin \alpha \, (b / r_0).
\ee

Expressing the time derivatives in equations (\ref{eq:p_phi})--(\ref{eq:E}) 
using equations (\ref{eq:r_dot})--(\ref{eq:phi_dot}) yields
\be
p_{\phi} = mvb \sin \theta_0 \sin \alpha
\ee
\be
\beta = m^2 v^2 b^2 + 2 m q p \cos \theta_0
\ee
\be
E = mv^2/2.
\ee
Substituting these results in equations (\ref{eq:beta}) and (\ref{eq:E}), 
we find
\be
\label{eq:r_dot2}
\dot{r}^2 = v^2 \left( 1 - \frac{b^2}{r^2} \right) - \frac{2q}{mr} \left( Q
+ \frac{p \cos \theta_0}{r} \right)
\ee
\be
\label{eq:theta_dot2}
r^4 \dot{\theta}^2 = v^2 b^2 \left( 1 - \frac{\sin^2 \theta_0 \sin^2 \alpha}
{\sin^2 \theta} \right) + \frac{2qp}{m} \left( \cos \theta_0 - \cos \theta
\right).
\ee 
We are interested in the solution with $\dot{r} < 0$, since the particle
approaches the grain.  The choice of the initial sign $S$ is more complicated 
for $\dot{\theta}$.  
If $\theta_0 = 0$ ($\pi$), then $S=+1$ ($S=-1$).  Otherwise, equation 
(\ref{eq:theta_dot}) yields 
$\dot{\theta} = v b \cos \alpha / r^2$ for $r \rightarrow \infty$.
Thus, $S=+1$ ($S=-1$) when $\cos \alpha > 0$ ($\cos \alpha < 0$).  
When $\cos \alpha = 0$, it is necessary to consider the second order 
term in the expansion for $\dot{\theta}$: 
$\sin \theta \, r \dot{\theta} = v \cos \theta_0 \, (b/r_0)^2$.
Thus, in this case, $S=\cos \theta_0/|\cos \theta_0|$.  If $\cos \theta_0$
and $\cos \alpha$ both equal zero, then $\dot{\theta} \equiv 0$.  
Note that $S$ typically changes sign at points $\theta$ where 
$\dot{\theta} = 0$.  

Since $\dot{r}^2$ is a single-valued function of $r$ (eq. \ref{eq:r_dot2}), 
charge $q$ only reaches the grain surface if $\dot{r}$ does not reverse 
sign when $r>a$.  From equation (\ref{eq:r_dot2}), $\dot{r} = 0$ when 
\be
\label{eq:r_turning}
r = a \left[ U \pm \sqrt{U^2
+ V \cos \theta_0 + \left( \frac{b}{a} \right)^2}
\right]
\ee
where $U \equiv qQ/(mv^2 a)$ and $V \equiv 2qp / (mv^2 a^2)$.
If the larger root in equation (\ref{eq:r_turning}), $r_+$, exceeds $a$
and $\ddot{r} > 0$ at $r = r_+$, then charge $q$ does not strike the grain.
If $U > 1$, then charge $q$ only strikes the grain if the 
argument of the square root in equation (\ref{eq:r_turning}) is negative, 
since $\dot{r}$ never reaches zero in this case.  Assuming $\ddot{r} > 0$
at $r_+$, the critical impact parameter is given by
\be
\label{eq:b_crit}
\frac{b_{\mathrm{crit}}}{a} = \cases{(1 - 2 U - V  
\cos \theta_0)^{1/2} &, $U \le 1$\cr
( -U^2 - V \cos \theta_0)^{1/2} &, $U \ge 1$\cr}.
\ee
Only trajectories with $b \le b_{\mathrm{crit}}$ strike the grain surface.
If the relevant root in equation (\ref{eq:b_crit}) is not real, then 
$b_{\mathrm{crit}} = 0$.   
Note that $b_{\mathrm{crit}}$ does not depend on the angle $\alpha$.  
The collision cross section is $\pi b_{\mathrm{crit}}^2$.  

Now we justify the assumption that $\ddot{r} > 0$ at $r = r_+$ when 
$b = b_{\mathrm{crit}}$.  Differentiating equation (\ref{eq:E}) yields
\be
\label{eq:diff_E}
\dot{r} \left( \ddot{r} - \frac{qQ}{mr^2} - \frac{\beta}{m^2 r^3} \right) 
= 0.
\ee
The term in parentheses in equation (\ref{eq:diff_E}) must vanish for all
$r$, except where $\dot{r} = 0$.  Continuity implies that it vanishes at
these locations as well, including at $r = r_+$.  Equations 
(\ref{eq:r_turning}) and (\ref{eq:diff_E}) yield
\be
\label{eq:ddot_r_turning}
\ddot{r} = \frac{a^2 v^2}{r_+^3} \left[ U^2 +  V \cos \theta_0 
+ \left( \frac{b}{a} \right)^2 + 
U \sqrt{ U^2 + V \cos \theta_0  + \left( \frac{b}{a} \right)^2} \right]
\ee
at $r = r_+$.  If either $U > 0$ or 
$V \cos \theta_0 > 0$, then clearly $\ddot{r} > 0$ at $r = r_+$, 
regardless of $b$.  If both of these quantities are negative, then setting
$b = b_{\mathrm{crit}}$ in equation (\ref{eq:ddot_r_turning}) yields
$\ddot{r} = |U-1| > 0$ for $r = r_+$.  

Assuming no gas-grain drift, the mean collision cross section
(averaged over angle $\theta_0$) is
\be
\bar{\sigma} = \cases{
\pi a^2 \left( 1 - 2 U \right) &, $2 U + |V| \le 1$\cr
0 &, $2 U - |V| \ge 1$\cr
\frac{1}{4} \pi a^2 |V|^{-1} ( 1 - 2 U + |V|)^2 &, otherwise\cr}
\ee
when $U \le 1$ and
\be
\bar{\sigma} = \cases{
\frac{1}{4} \pi a^2 |V| ( 1 - |V|^{-1} U^2)^2 &, 
$|V|^{-1} U^2 < 1$\cr
0 &, $|V|^{-1} U^2 \ge 1$\cr}
\ee
when $U \ge 1$.  

\begin{figure}
\includegraphics[width=84mm]{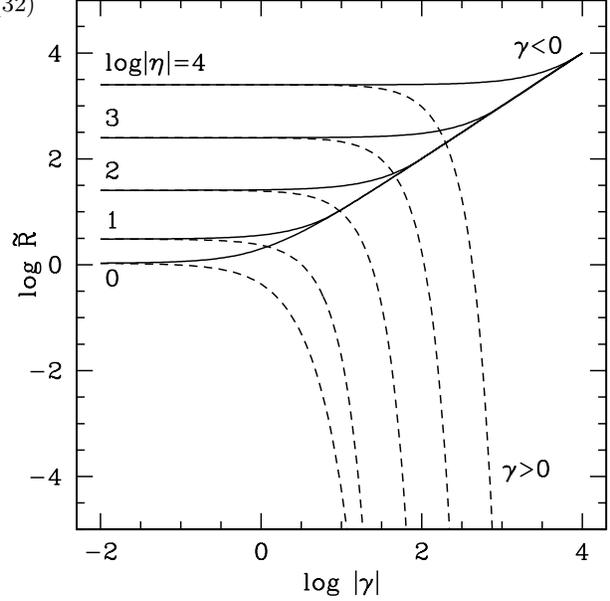}
\caption{
$\tilde{R}$ vs. $\gamma$ for various values of $|\eta|$, as indicated.
        }
\label{fig:R_tilde}
\end{figure}

Integrating over the Maxwell speed distribution yields the factor $\tilde{R}$
from equation (\ref{eq:R_coll}) 
for a grain that does not drift relative to the gas:
\be
\label{eq:R_tilde1a}
\tilde{R}(\gamma, |\eta|) = 1 - \gamma ,~~qQ < 0~~\mathrm{and}~~|\eta| 
\le - \gamma
\ee
\begin{eqnarray}
\label{eq:R_tilde1b}
\nonumber
\tilde{R} & = &  
\frac{1}{4 |\eta|} \left\{ (|\eta| - \gamma) (2 + |\eta| - \gamma) + 2 \left[
1 - \mathrm{e}^{-(\gamma + |\eta|)} \right] 
\right\} ,\\
& & qQ < 0~~\mathrm{and}~~|\eta| 
\ge - \gamma
\end{eqnarray}
\be
\label{eq:R_tilde2a}
\tilde{R} = \mathrm{e}^{- \gamma} \, \frac{\sinh |\eta|}{|\eta|}
,~~qQ > 0~~\mathrm{and}~~|\eta| \le \gamma / 2
\ee
\begin{eqnarray}
\label{eq:R_tilde2b}
\nonumber
\tilde{R} & = & - \frac{1}{2 |\eta|} \mathrm{e}^{-(\gamma + |\eta|)}\\
\nonumber
& & + \frac{|\eta|^2 + (2- \gamma) |\eta|
+ (2 - \gamma + \gamma^2/4)}{4 |\eta|} \mathrm{e}^{-\gamma/2}\\
\nonumber
& & + \frac{|\eta|}{4}
\int_{\gamma^2/(4 |\eta|)}^{\gamma/2} ds \left( 1 - \frac{\gamma^2}{4 |\eta| s}
\right)^2 \mathrm{e}^{-s},\\
& & qQ > 0~~\mathrm{and}~~|\eta| \ge \gamma / 2 ;
\end{eqnarray}
$\gamma \equiv qQ/(a k_{\mathrm{B}} T_{\mathrm{gas}})$, and 
$\eta \equiv q p / (a^2 k_{\mathrm{B}} T_{\mathrm{gas}})$.
Fig. \ref{fig:R_tilde} displays $\tilde{R}$ versus $\gamma$ for various
values of $|\eta|$.  Note that equations (\ref{eq:R_tilde1a}) and 
(\ref{eq:R_tilde2a}) recover the classic Spitzer (1941) expression for 
$\tilde{R}$ for a charged sphere when $\eta = 0$.

Since $\mathrm{d}\theta/\mathrm{d}r = \dot{\theta}/\dot{r}$, equations 
(\ref{eq:r_dot2}) and (\ref{eq:theta_dot2}) yield
\be
\label{eq:F1F2}
F_1 \left( \theta_0, \theta, VC^2, \alpha 
\right) = F_2(A, B,C)
\ee
where 
\be
A \equiv 1 + V C^2 \cos \theta_0
\ee
\be
B \equiv 2 U C
\ee
\be
C \equiv a/b
\ee
\begin{equation}
\label{eq:F_1}
F_1 \left( \theta_0, \theta, V C^2, \alpha 
\right) = 
\end{equation}
\begin{displaymath}
\int_{\theta_0}^{\theta} \frac{|\sin \theta^{\prime}|
\mathrm{d} \theta^{\prime}
S(\theta^{\prime})}
{\sqrt{\sin^2 \theta^{\prime} - \sin^2 \theta_0 \sin^2 \alpha + 
V C^2 \sin^2 \theta^{\prime} 
(\cos \theta_0 - \cos \theta^{\prime})}}
\end{displaymath}
\be
\label{eq:F_2a}
F_2 (A, B, C) = \int_C^{\infty} \frac{du}{u \sqrt{-A - Bu + u^2}}.
\ee
The integrand in equation (\ref{eq:F_1}) is negative when $S < 0$, but 
in these cases $\theta < \theta_0$, so the integral remains positive.
If $\dot{\theta}$ reaches zero at $\theta^{\prime} = \theta_1$, then
$S$ changes sign and $F_1$ splits into two integrals, with limits 
$\theta_0$ to $\theta_1$ and $\theta_1$ to $\theta$.  
Performing the integration in equation (\ref{eq:F_2a}), 
\begin{eqnarray}
\nonumber
F_2 & = &  
\frac{1}{\sqrt{A}} \left[ \sin^{-1} \left( \frac{BC + 2A}{C \sqrt{B^2 + 
4A}} \right)
- \sin^{-1} \left( \frac{B}{\sqrt{B^2 + 4A}} \right) \right] ,\\
& & A \ge 0
\end{eqnarray}
\begin{eqnarray}
\nonumber
F_2 & = &
\frac{1}{\sqrt{-A}} \ln \left[ \frac{2 \sqrt{-A} \sqrt{C^2 - BC -A} - BC
-2A}{(2 \sqrt{-A} - B) C} \right] ,\\
& & A \le 0.
\end{eqnarray}

Given $2qp/(mv^2b^2)$, $2qQ/(mv^2 b)$, $a/b$, $\theta_0$, and $\alpha$,
equation (\ref{eq:F1F2}) can be solved to efficiently find the arrival angle
$\theta$.  A less efficient, but more direct, approach is to integrate the
equations of motion (\ref{eq:r_dot_dot})--(\ref{eq:phi_dot_dot}).
We have written {\sc fortran}
subroutines implementing both of these methods and 
found perfect agreement for numerous combinations of input parameters.  

To compute the distribution of arrival angles 
$\theta$ for given values of $\gamma$ and $\eta$, we examine a large 
number of trajectories with initial parameters $\theta_0$, 
$u \equiv v/v_{\mathrm{th}}$, $b/a$, and $\alpha$, where 
$v_{\mathrm{th}} \equiv (2 k_{\mathrm{B}} T_{\mathrm{gas}}/m)^{1/2}$.
We first select $N_{\theta}$ values of $\theta_0$ from 0 to $\pi$, uniformly 
spaced in $\cos \theta_0$.  For each value of $\theta_0$, we select 
$N_v$ values of $u$, starting with $u = 1.08765$, 
the median value assuming the Maxwell speed distribution.  
We then select $(N_v - 1)/2$ values with $u > 1.08765$ spaced in 
equal-probability intervals, i.e., such that
\be
\label{eq:u_spacing}
\frac{4}{\sqrt{\pi}} \int_{u_i}^{u_{i+1}} \mathrm{d} u \, 
u^2 \exp(-u^2) = \frac{1} {N_v+1}~~~.
\ee
Likewise for values with $u < 1.08765$.
For each $(\theta_0, u)$ pair, if $b_{\mathrm{crit}} > 0$, then we next select
$N_b$ values of $b/a$ between 0 and $b_{\mathrm{crit}}/a$, uniformly spaced in 
$b^2$.  Finally, for each $(\theta_0, u, b/a)$, we select $N_{\alpha}$ 
values of $\alpha$, uniformly spaced between 0 and $2\pi$.
For each trajectory, we compute the arrival angle $\theta$.  The results
are binned, with trajectories weighted in proportion to $b_{\mathrm{crit}}^2$.

\begin{figure}
\includegraphics[width=84mm]{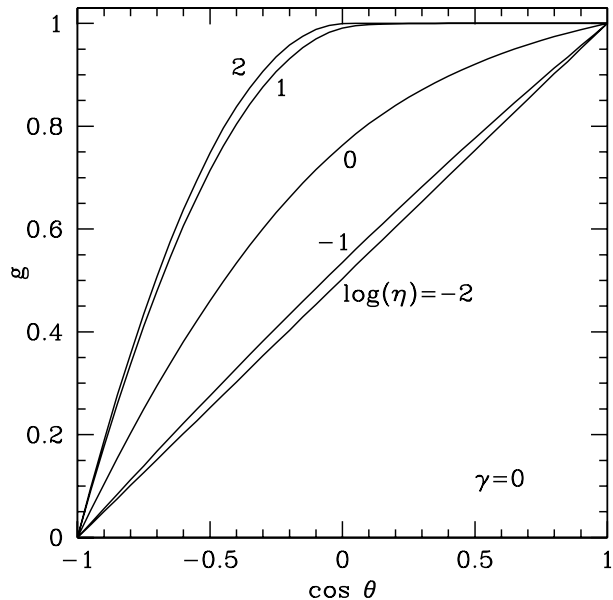}
\caption{
$g(\cos \theta)$ vs. $\cos \theta$ for $\gamma = 0$ and various values of 
$\eta$, as indicated.
        }
\label{fig:g_gamma0}
\end{figure}

\begin{figure}
\includegraphics[width=84mm]{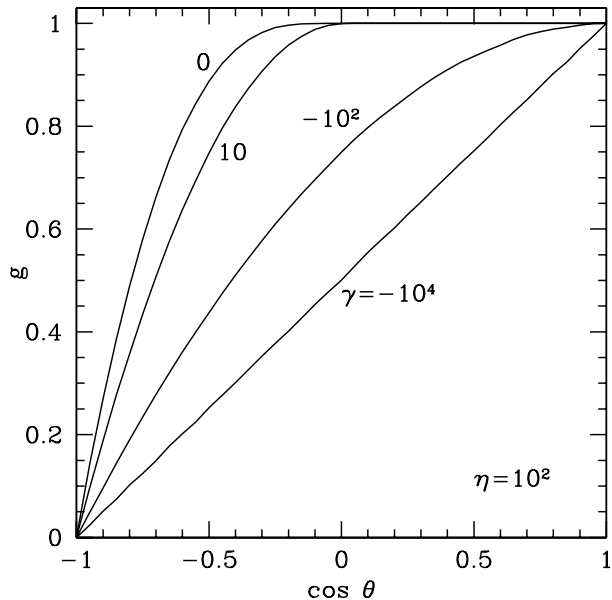}
\caption{
$g(\cos \theta)$ vs. $\cos \theta$ for $\eta = 10^2$ and various values of 
$\gamma$, as indicated.
        }
\label{fig:g_eta2}
\end{figure}

Fig. \ref{fig:g_gamma0} displays $g(\cos \theta)$, the fraction of 
arriving particles that strike with cosine of the polar angle (relative to 
the dipole
moment) $\le \cos \theta$, for $\gamma =0$ and several values of $\eta$.  
To construct this figure, we adopted 40 bins in $\theta$ and
$N_{\theta} = N_v = N_b = N_{\alpha} = 31$.  The distributions look very
similar to those in fig. \ref{fig:g_gamma0} when $|\gamma| < 1$.  
Distributions for $(\gamma, -\eta)$ are identical to those for 
$(\gamma, \eta)$, except that they are referenced to $\cos \theta = 1$
rather than -1.  That is, 
$g(\gamma, -\eta; \cos \theta) = g(\gamma, \eta; - \cos \theta)$, 
with $g(\cos \theta)$ the fraction of particles that strike with 
cosine of polar angle $\ge \cos \theta$ when $\eta < 0$.   
As $|\gamma|$ increases, the
distribution in $\cos \theta$ becomes more uniform, as seen in Fig.
\ref{fig:g_eta2} for the case that $\eta = 10^2$.

Electrons arriving at the grain surface can penetrate to within the bulk of
the grain, with an e-folding length $l_e \sim 10 \ \mathrm{\AA}$ 
(see paragraph
following eq. 13 in WD01).  We neglect this penetration since $l_e \ll a$;
i.e., all arriving electrons are assumed to be located at $r=a$.

\subsection{Photoelectric Emission}

We adopt a simplified version of the procedure in WD01 for calculating
the rate at which photoelectrons are ejected from the grain, $J_{\mathrm{pe}}$.
WD01 express the photoelectric yield (i.e., the probability that an electron
is ejected following the absorption of a photon) as a product of three 
factors: the bulk yield $y_0$, a size-dependent yield enhancement factor 
$y_1$, and a term $y_2$ that accounts for the attraction of `attempting'
photoelectrons back to the grain when $Z \ge 0$.  (Recall that the grain 
charge $Q = Z e$, with $e$ the proton charge.)
For the relatively large
grains under consideration here, $y_1 = 1$.  The term $y_2$ is given by
(WD01, eq.~11)
\be
y_2 = \cases{E_{\mathrm{high}}^2 (E_{\mathrm{high}} - 3 E_{\mathrm{low}}) / 
(E_{\mathrm{high}} - E_{\mathrm{low}})^3  &, $Z \ge 0$\cr
1 &, $Z < 0$\cr} ,
\ee
with $E_{\mathrm{low}} = - (Z+1) e^2 /a$ and 
$E_{\mathrm{high}} = h\nu - h\nu_{\mathrm{pet}}$ ($h\nu_{\mathrm{pet}}$ is the 
threshold photon energy for photoemission).  For simplicity, we take 
$E_{\mathrm{high}} = 3 \ \mathrm{eV}$ [assuming $h\nu \approx 11 \ \mathrm{eV}$ 
and 
$h\nu_{\mathrm{pet}} \approx 8 \ \mathrm{eV}$ (WD01)], independent of $Z$ and 
$h\nu_{\mathrm{pet}}$.  

To find $J_{\mathrm{pe}}$, it is necessary to integrate the photon absorption 
rate $R_{\gamma}$ 
times the yield over the range of available photon energies above 
$h\nu_{\mathrm{pet}}$.  Since we approximate $y_2$ to be independent of $h\nu$,
$J_{\mathrm{pe}} \propto y_2$.  We simply choose the proportionality constant
so as to reproduce the average grain potential of 
$\approx 0.3 \ \mathrm{V}$ from WD01.

As for arriving electrons, we assume that holes produced in photemission
events are located at $r=a$.  We also neglect the production of an 
electron-hole pair when a photon is absorbed but a photoelectron does not
escape the grain.  In some cases, the photon absorption occurs too deep
within the grain for the photoelectron to reach the surface or the 
photoelectron's velocity is directed away from the surface (resulting in 
$y_0 < 1$).  Since $l_e \ll a$, the resulting separation of charge does not 
contribute significantly to the dipole moment.  In other cases, a 
photoelectron breaches the grain surface, but returns to the grain due to
an attractive Coulomb force if $Z \ge 0$ (resulting in $y_2 < 1$).  Such 
events could lead to a more significant change in $\mathbf{p}$, but are 
rare; $y_2 \approx 0.98$ when the grain potential is $0.3 \ \mathrm{V}$.

\section{SIMULATIONS}
\label{sec:simulations}

For each of the four charge transport models described in \S 
\ref{sec:idealizations}, we run stochastic simulations that keep track of the 
grain dipole moment $\bmath{p}$ and the orientation in space of the grain's
rotational axis, assumed fixed with respect to the grain body (as would be 
appropriate for suprathermally rotating grains).  We adopt 
$v_{\perp} = 1 \ \mathrm{km} \ \mathrm{s}^{-1}$ (resulting from acceleration 
associated with magnetohydrodynamic turbulence; Yan et al.~2004),
$T_{\mathrm{gas}} = 100 \ \mathrm{K}$, 
$n_e = 4.5 \times 10^{-2} \ \mathrm{cm}^{-3}$, $s_e = 0.5$, 
$T_{\mathrm{d}} = 15 \ \mathrm{K}$ (hence, $\chi_0 = 5 \times 10^{-3}$), and 
$y_0 = 6 \times 10^{-2}$ (eq. 17 in WD01 with $h\nu = 10 \ \mathrm{eV}$).  
We consider grains with $a=0.1$ and $0.2 \ \micron$, for which the photon 
absorption rate $R_{\gamma} = 2.9 \times 10^{-2}$ and 
$5.7 \times 10^{-2} \ \mathrm{s}^{-1}$, respectively, in order to maintain the 
average potential at $0.3 \ \mathrm{V}$.  We employ a constant time step size 
$\mathrm{d}t$, usually $31.56 \ \mathrm{s}$, which is smaller than the 
typical time between charging events.  

Next, we describe the simulation algorithm for perfectly insulating grains.
At the start of each time step, we find the factor $\tilde{R}$ 
(eq. \ref{eq:R_coll}) for the electron collisional charging rate by bilinear
interpolation (Press et al. 1992, p. 117) in $\ln \gamma$ and $\ln \eta$, 
with 21 values of $\eta$ ranging from $10^{-2}$ to $10^2$ and 11 values of
$|\gamma|$ ranging from $10^{-2}$ to $3 \times 10^2$ (for 23 total values 
of $\gamma$ in the tables, since both signs, as well as $\gamma=0$, are
included).  If $|\gamma| < 10^{-2}$, then a linear interpolation is 
performed in $\eta$ alone (with $\gamma =0$).  If $\eta < 10^{-2}$, then we 
assume the classic Spitzer (1941) expression for $\tilde{R}$ for a charged 
sphere (corresponding to $\eta = 0$).  

An electron arrives with probability $R_e \, \mathrm{d}t$ 
($R_e$ is the electron arrival rate; eq. \ref{eq:R_coll}).  
In each time-step, we check that  $R_e \, \mathrm{d}t < 1$ (and likewise for
the probability that a photoelectron is ejected).  
Here, as throughout the simulations, we 
use the routine {\sc ran2} from Press et al. (1992) for choosing random 
numbers.  If $\eta < 10^{-2}$, then the electron is placed at a random 
location $(\theta, \phi$) on the grain surface; $\theta$ and $\phi$ are the 
polar and azimuthal angles, respectively, with the rotation axis 
$\bmath{\hat{\omega}} = \bmath{\hat{z}}$ as the polar axis.
Otherwise, we interpolate to find the distribution function 
$g(\theta^{\prime})$, as described in the preceding paragraph for $\tilde{R}$;
$\theta^{\prime}$ is the polar angle with the dipole moment $\bmath{p}$ as
the polar axis.  We choose $\theta^{\prime}$ randomly from the distribution
$g(\theta^{\prime})$ and the azimuthal angle $\phi^{\prime}$ is chosen
randomly from a uniform distribution.  The arrival position with respect to
the grain body is given by
\begin{eqnarray}
\nonumber
x/a & = & \sin \theta^{\prime} (\cos \phi^{\prime} \cos \theta_p \cos \phi_p - 
\sin \phi^{\prime} \sin \phi_p)\\
& & + \cos \theta^{\prime} \sin \theta_p 
\cos \phi_p
\end{eqnarray}
\begin{eqnarray}
\nonumber
y/a & = & \sin \theta^{\prime} (\cos \phi^{\prime} \cos \theta_p \sin \phi_p + 
\sin \phi^{\prime} \cos \phi_p)\\
& & + \cos \theta^{\prime} \sin \theta_p 
\sin \phi_p
\end{eqnarray}
\be
z/a = \cos \theta^{\prime} \cos \theta_p - \sin \theta^{\prime} \cos 
\phi^{\prime} \sin \theta_p 
\ee
where $\theta_p$ and $\phi_p$ are the polar and azimuthal angles, 
respectively, of $\bmath{p}$ relative to the grain body.  

A photoelectron is ejected with probability $R_{\gamma} \, y_0 \, y_2$.
The resulting hole is located randomly on the grain surface.  

At the start of a simulation, we set $\bmath{p} = 0$ and choose $Z$ to 
correspond to the average potential of $0.3 \ \mathrm{V}$.  
Draine \& Lazarian (1998) noted that a grain may have an intrinsic electric
dipole moment due to the random orientations of polar constituents.  Thus,
our choice $\bmath{p} = 0$ requires justification.  

Consider a conducting grain with $N_{\mathrm{dt}}$ deep traps (model 3
in \S \ref{sec:idealizations}).  The largest possible magnitude of the
electric dipole moment associated with excess charges (electrons and holes)
occupying the deep traps is $p_{\mathrm{max}} \sim N_{\mathrm{dt}} \, e a$.
If the magnitude of the intrinsic electric dipole moment
$p_{\mathrm{int}} > p_{\mathrm{max}}$, then the intrinsic dipole could not
be neutralized; thus, flipping of $\bmath{p}$ would not be possible.  This
situation is similar to that of a purely conducting grain (model 2 in
\S \ref{sec:idealizations}) in the cold neutral medium, for which
$p \propto Z$ and $Z$ is always positive.

To estimate the
likely magnitude of $p/ea$ associated with the intrinsic dipole, suppose 
each polar constituent has volume $V_0$ and dipole moment 
$p_0 = \zeta e V_0^{1/3}$.  Assuming each constituent is randomly oriented,
the total instrinsic moment $p_{\mathrm{int}} \sim N^{1/2} p_0$, with the 
grain volume $(4/3) \pi a^3 = N V_0$.  Eliminating $N$, we find
\be
\frac{p_{\mathrm{int}}}{ea} \sim \left( \frac{4 \pi}{3} \right)^{1/2} \zeta
\left( \frac{a}{V_0^{1/3}} \right)^{1/2}.
\ee
Even adopting relatively large values of $\zeta \sim 0.1$ and 
$a V_0^{-1/3} \sim 500$, we find $p_{\mathrm{int}}/ea \sim 4.6$, comparable 
(in order of magnitude) to
the values found in the following section, where the instrinsic electric 
dipole moment $\bmath{p_{\mathrm{int}}}$ is neglected.  

For any realistic grain, $N_{\mathrm{dt}} \gg p_{\mathrm{int}}/ea$; thus,
we do not expect the intrinsic electric dipole moment to play any role in
the long-term evolution of $\bmath{p}$, including the flipping of the
dipole moment.  Essentially, the total number of charges in the grain (the
number of electrons plus the number of holes) can vastly exceed the net
number of charges (number of electrons minus number of holes), and a slight
asymmetry in the distribution of these charges can counter the intrinsic
dipole moment.

Of all the simulations with deep traps considered in this paper, the
smallest value of $N_{\mathrm{dt}}$ is $\approx 100$, when $a = 0.1 \micron$
and the volume per deep trap is $V_t = 4 \times 10^7 \mathrm{\AA}^3$.  
(This value of $N_{\mathrm{dt}}$ is almost certainly much too small to be 
realistic, but was chosen to make the computations feasible and to, 
conservatively, generate a grain with low insulating capability).  
We ran this model 3 simulation with $p_{\mathrm{int}} = 0$ and with
$p_{\mathrm{int}}$ as estimated above; the resulting 
flipping times are nearly identical, as expected.

At any time, the net charge and dipole moment are given by
\be
Z = N_h - N_e 
\ee
\be
\bmath{p} = e \sum_{i=1}^{N_h} \bmath{x}_i - e \sum_{i=1}^{N_e} \bmath{x}_i
\ee
where $N_h$ and $N_e$ are the total number of holes and electrons, 
respectively, and $\bmath{x}_i$ is the position of an electron or hole
(with the origin at the grain's center of mass, i.e., the center of the 
spherical grain).  
In each time-step, $\theta_p$, $\phi_p$ and $p \equiv |\bmath{p}|$ are
updated, if an electron arrives at or departs the grain.  
We also keep track of $\theta_{\mathrm{align}}$ and
$\phi_{\mathrm{align}}$, the polar and azimuthal angles of the grain rotation 
axis with respect to the magnetic field direction, employing eqs. 14 and 15
from W06:
\be
\label{eq:dphi_align}
\mathrm{d}\phi_{\mathrm{align}} = \Omega_0 [ 1 - \Upsilon \cot 
\theta_{\mathrm{align}} \cos ( \phi_{\mathrm{align}} + \phi_{\mathrm{gyro}})
] \, \mathrm{d}t 
\ee 
\be
\label{eq:dtheta_align}
\mathrm{d}\theta_{\mathrm{align}} = - \Omega_0 \Upsilon \sin ( 
\phi_{\mathrm{align}} + \phi_{\mathrm{gyro}} ) \, \mathrm{d}t
\ee
where $\Omega_0$ is the precession rate for the case that $\bmath{p} =0$
(eq. \ref{eq:prec}),
\be
\phi_{\mathrm{gyro}}(t) = \int_0^t \mathrm{d}t^{\prime} \, 
\omega_{\mathrm{gyro}}(t^{\prime}) ,
\ee 
and the time-scale for gyrorotation is given by
\begin{eqnarray}
\nonumber
\omega_{\mathrm{gyro}}^{-1} & \sim & 2.4 \times 10^2 \left( \frac{\rho}{3 \ 
\mathrm{g} \ \mathrm{cm}^{-3}} \right) \left( \frac{a}{0.1 \ \micron} 
\right)^2 \left( \frac{U}{0.3 \ \mathrm{V}} \right)^{-1}\\
& & \times \left( \frac{B}
{5 \ \mu \mathrm{G}} \right)^{-1} \ \mathrm{yr}.
\end{eqnarray}
Note that $\omega_{\mathrm{gyro}}$ varies with time, since the grain 
potential $U$ is not constant. 

We take $\theta_{\mathrm{align}} = 0.1$ and $\phi_{\mathrm{align}} = 0$ initially.
Within the same charging simulation, we consider several different values
of $\omega / \omega_T$ (and thus, several different values of $\Upsilon$; 
recall eqs. \ref{eq:omega_T}, \ref{eq:mu_Bar_sil}, and \ref{eq:Upsilon}).
In principle, gyrorotation can affect the disalignment, since 
$\omega_{\mathrm{gyro}}$ fluctuates randomly as $Z$ does so.  However, we have
found that the disalignment time is identical for simulations that do
(do not) include gyrorotation.  Thus, we omit gyrorotation in our 
simulations.  

The simulations for purely conducting grains are identical to those for
purely insulating grains, except that it is not necessary to keep track of
the electron arrival locations, since charge is immediately delocalized.
Instead, we simply take $\bmath{p} = \bmath{p_z \hat{z}} \propto Z$.   

For the models containing deep traps, we first specify the average grain 
volume per deep trap of a given type (i.e., a trap that accomodates electrons
versus one that accomodates holes), $V_t$, then randomly place 
${\rm int}(4\pi a^3 / 3 V_t)$ deep traps of each type throughout the grain
 volume.  
For model 3 in \S \ref{sec:idealizations} (conducting grain with deep traps), 
an arriving electron is immediately moved to the accomodating trap nearest 
its arrival site.  This nearest trap could be a vacant electron
trap or an occupied hole trap; in the latter case, the charges recombine.
Likewise, the hole produced in a photoemission event is immediately moved
to the nearest vacant hole trap or occupied electron trap.

For model 4 in \S \ref{sec:idealizations} (partially conducting grain with 
deep traps), each electron or hole undergoes a random walk through the grain,
starting at its arrival location.  In each step, the charge moves distance
$d_{\mathrm{rw}}$ (taken to be $30 \ \mathrm{\AA}$) in time 
$t_{\mathrm{rw}}$.  Thus, for these simulations, the time step size 
$\mathrm{d}t = t_{\mathrm{rw}}$.  If a charge finds itself within distance 
$d_{\mathrm{rw}}$ of an accomodating deep trap, then it enters the trap and 
remains there until recombining when a charge with opposite sign arrives at 
the trap.  The time $t_{\mathrm{rw}}$ is selected as follows:
\be
t_{\mathrm{rw}} = \frac{f \tau_c d_{\mathrm{rw}}^2}{V_t^{2/3}}
\ee
with $\tau_c$ the typical time between charging events.  
With this choice, the typical time for a charge to travel from one trap to
another is $\sim f \tau_c$.  With $f \sim 1$, this model lies between the
extremes of a perfect insulator and a perfect conductor with deep traps.

\section{RESULTS}
\label{sec:results}

We ran simulations for 11 different sets of input, with 2 realizations apiece
(i.e., 2 different values of the random number seed), for a total of 22
simulations.  Table \ref{tab:sim} displays input parameters for each run, 
as well as selected output parameters.  We performed runs with models 1 through
3 of \S \ref{sec:idealizations} for grain radii $a = 0.1$ and 
$0.2 \ \micron$.  For model 4, only $a = 0.1 \ \micron$ is included, since 
the CPU time becomes prohibitive for $a = 0.2 \ \micron$ when charges 
execute random walks through the grain volume.  For models 1 through 3,
the total duration of the simulation is 
$t_{\mathrm{tot}} = 10^5 \ \mathrm{yr}$, but 
substantially shorter $t_{\mathrm{tot}}$ were obtained for model 4.

\begin{table*}
\centering
\begin{minipage}{140mm}
\caption{Simulation Parameters and Outputs \label{tab:sim}}
\begin{tabular}{@{}lllllllllll@{}}
\hline
 & & & & & & & & \multicolumn{3}{c}{$N_{\mathrm{dev}}$\footnote{Number of
$1 \ \mathrm{rad}$ deviations in alignment angle $\theta_{\mathrm{align}}$.}}\\
Model\footnote{From section \ref{sec:idealizations}.} &
$a$\footnote{Grain radius.} &
$V_t$\footnote{Volume per deep trap.} &
$t_{\mathrm{rw}}$\footnote{Duration of random walk step.} &
Run &
$|p_z|_{\mathrm{av}}$\footnote{Average of the absolute value of the component 
of the 
electric dipole moment lying along the spin axis (normalized to $ea$, the
product of the proton charge and the grain radius).} &
$\tau_{\mathrm{flip}}$\footnote{Estimate of the electric dipole moment flipping 
time.} &
$t_{\mathrm{tot}}$\footnote{Duration of the simulation.} &
1.0\footnote{Suprathermality $\omega / \omega_T$.} &
1.5$^i$ &
2.0$^i$\\
  &
$\micron$ &
$\mathrm{\AA}^3$ &
s &
  &
$ea$ &
$10^{-4} \,$yr &
yr &
  &
  &
 \\
\hline
1 &0.1 &...   &...  &1 &2.21 &5.2 &1.0E5  &2348 &218 &30\\
1 &0.1 &...   &...  &2 &2.18 &5.2 &1.0E5  &2418 &195 &26\\
2 &0.1 &...   &...  &1 &2.11 &...    &1.0E5  &72   &5   &1\\
2 &0.1 &...   &...  &2 &2.12 &...    &1.0E5  &44   &6   &2\\
3 &0.1 &4.0E7 &...  &1 &1.77 &6.8 &1.0E5  &1822 &179 &19\\
3 &0.1 &4.0E7 &...  &2 &1.77 &6.2 &1.0E5  &1800 &160 &19\\
3 &0.1 &4.0E6 &...  &1 &2.11 &5.8 &1.0E5  &2451 &220 &26\\
3 &0.1 &4.0E6 &...  &2 &2.13 &5.8 &1.0E5  &2358 &228 &32\\
4 &0.1 &4.0E7 &2.3  &1 &2.16 &2.2 &7.29E2 &14   &1   &0\\
4 &0.1 &4.0E7 &2.3  &2 &2.35 &2.5 &7.29E2 &22   &1   &0\\
4 &0.1 &4.0E7 &1.15 &1 &2.17 &2.5 &3.64E2 &6    &1   &0\\
4 &0.1 &4.0E7 &1.15 &2 &2.09 &2.4 &3.64E2 &11   &1   &0\\
4 &0.1 &4.0E7 &9.2  &1 &2.56 &2.4 &2.92E3 &81   &11  &1\\
4 &0.1 &4.0E7 &9.2  &2 &2.56 &2.5 &2.92E3 &76   &7   &1\\
4 &0.1 &4.0E6 &2.3  &1 &2.07 &5.0 &7.29E2 &17   &2   &0\\
4 &0.1 &4.0E6 &2.3  &2 &2.11 &5.1 &7.29E2 &19   &3   &0\\
1 &0.2 &...   &...  &1 &2.23 &2.6 &1.0E5  &172  &16  &2\\
1 &0.2 &...   &...  &2 &2.18 &2.6 &1.0E5  &150  &12  &1\\
2 &0.2 &...   &...  &1 &2.00 &...    &1.0E5  &4    &0   &0\\
2 &0.2 &...   &...  &2 &2.00 &...    &1.0E5  &4    &1   &0\\
3 &0.2 &4.0E7 &...  &1 &2.07 &2.9 &1.0E5  &164  &17  &2\\
3 &0.2 &4.0E7 &...  &2 &2.07 &2.9 &1.0E5  &139  &14  &2\\
\hline
\end{tabular}
\end{minipage}
\end{table*}

\begin{figure}
\includegraphics[width=84mm]{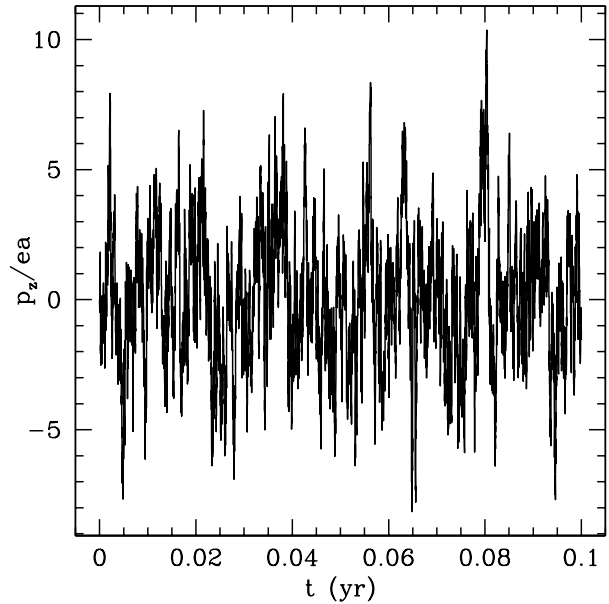}
\caption{
The component of the grain dipole moment (normalized to $ea$, the proton charge
times the grain radius) lying along the spin axis vs. time,
from a simulation of a purely insulating grain with $a=0.1 \ \micron$.  
        }
\label{fig:p_z}
\end{figure}

Fig. \ref{fig:p_z} displays the component of the electric dipole moment
lying along the spin axis, $p_z$, from a simulation for a purely insulating
grain with $a=0.1 \ \micron$.  Clearly, $p_z$ reverses sign on a short 
time-scale of a fraction of a year.  We estimate that the dipole flipping 
time-scale $\tau_{\mathrm{flip}} \approx t_{\mathrm{tot}} / N_{\mathrm{flip}}$, 
where $N_{\mathrm{flip}}$ is the total number of dipole flips that occur in the 
simulation.  We take a flip to occur each time $|p_z|$ increases beyond unity 
with $p_z$ having the opposite sign as it did the previous time $|p_z|$ 
increased past unity.  As seen in Table \ref{tab:sim}, 
$\tau_{\mathrm{flip}} < 10^{-3} \ \mathrm{yr}$ for all simulations.

The precession time $|\Omega_0|^{-1} \approx 2.7 \times 10^{-3} \ \mathrm{yr}$
($1.1 \times 10^{-2} \ \mathrm{yr}$) for grains with $a=0.1 \ \micron$ 
($0.2 \ \micron$).  Thus, it is always the case that 
$\tau_{\mathrm{flip}} < |\Omega_0|^{-1}$ (though not always that 
$\tau_{\mathrm{flip}} \ll |\Omega_0|^{-1}$).  This suggests that equation 
(\ref{eq:tau_dis1}), i.e. $\tau_{\mathrm{dis}} \sim \Upsilon^{-2} 
|\Omega_0|^{-2} \tau_{\mathrm{flip}}^{-1}$, may be a good approximation for 
the disalignment time $\tau_{\mathrm{dis}}$.  Since
$\Upsilon \propto \omega^{-1}$ (eqs. \ref{eq:mu_Bar_sil} and 
\ref{eq:Upsilon}), $\tau_{\mathrm{dis}} \propto (\omega / \omega_T)^2$.
(When $\Upsilon  \ll 1$, as it is for suprathermally rotating silicate
grains, $\tau_{\mathrm{dis}} \propto \Upsilon^{-2} \propto \omega^2$ when 
$\tau_{\mathrm{flip}} \gg |\Omega_0|^{-1}$ as well; eq. \ref{eq:tau_dis2}.)
As a result, extremely long simulation times are needed to estimate 
$\tau_{\mathrm{dis}}$ for suprathermally rotating grains.  However, this 
proportionality also suggests a strategy for obtaining useful information
with shorter simulations.  We run for several values of $\omega / \omega_T$,
as low as 0.1, and check the scaling of $\tau_{\mathrm{dis}}$ versus 
$\omega / \omega_T$.  In fact, we find that usually 
$\tau_{\mathrm{dis}} \propto (\omega / \omega_T)^2$; thus, 
$\tau_{\mathrm{dis}}$ for
highly suprathermally rotating grains can be estimated by extrapolation.
Note, also, that $\Upsilon^{-2} |\Omega_0|^{-2}$ is independent of the 
magnetic dipole moment $\mu$, and hence is independent of $\chi_0$.
Consequently, $\tau_{\mathrm{dis}}$ does not depend on this highly uncertain 
parameter in this regime.  

To estimate $\tau_{\mathrm{dis}}$ from a simulation, we keep track of 
$N_{\mathrm{dev}}$, the number of times that $\theta_{\mathrm{align}}$ suffers 
a deviation of $1 \ \mathrm{rad}$.  Once $\theta_{\mathrm{align}}$ reaches a 
value $\theta_{\mathrm{align}, 1}$ differing by $1 \ \mathrm{rad}$ from its 
initial value, $N_{\mathrm{dev}} = 1$. When it reaches a value differing by 
$1 \ \mathrm{rad}$ from $\theta_{\mathrm{align}, 1}$, $N_{\mathrm{dev}} = 2$, 
etc.  When $N_{\mathrm{dev}} \gg 1$, 
$\tau_{\mathrm{dis}} \approx t_{\mathrm{tot}} / N_{\mathrm{dev}}$.  For the 
relatively small values of $\omega / \omega_T$ under consideration, this 
condition obtains.  However, this is not typically the case when 
$\omega / \omega_T \ga 10$.  In general, we estimate 
$\tau_{\mathrm{dis}} \approx t_{\mathrm{tot}} / (N_{\mathrm{dev}} + 
|\Delta \theta_{\mathrm{align}}|)$, where $\Delta \theta_{\mathrm{align}}$ is the 
value of $\theta_{\mathrm{align}}$ at the end of the simulation minus its value 
at the last time $N_{\mathrm{dev}}$ was incremented (which may have been the 
start of the simulation, if $N_{\mathrm{dev}} = 0$).  Table \ref{tab:sim} 
indicates the values of $N_{\mathrm{dev}}$ for 
$\log_{10}(\omega / \omega_T) = 1.0$, 1.5, and 2.0.  Of course, the resulting 
estimate of $\tau_{\mathrm{dis}}$ is not 
very reliable for the cases where $N_{\mathrm{dev}} \sim 1$.

\begin{figure}
\includegraphics[width=84mm]{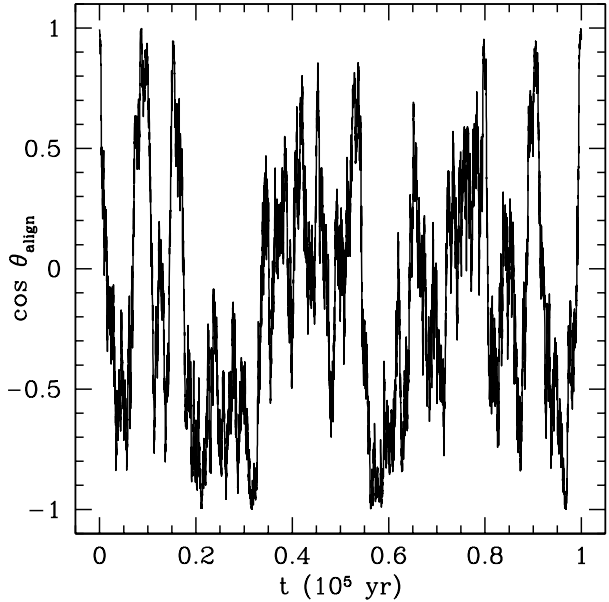}
\caption{
$\cos \theta_{\mathrm{align}}$ versus time from a simulation of a perfectly 
insulating grain with $a=0.1 \ \micron$ and $\omega / \omega_T = 10^2$;
$N_{\mathrm{dev}} = 30$ for this case.   
        }
\label{fig:cos_theta_align}
\end{figure}

For each simulation, we keep track of $\theta_{\mathrm{align}}$ for 7 different
values of $\log_{10} (\omega / \omega_T)$, evenly spaced between -1.0 and 2.0.  
Fig. \ref{fig:cos_theta_align} displays $\cos \theta_{\mathrm{align}}$ 
versus time from a simulation of a perfectly insulating grain with 
$a=0.1 \ \micron$ and $\omega / \omega_T = 10^2$.

\begin{figure}
\includegraphics[width=84mm]{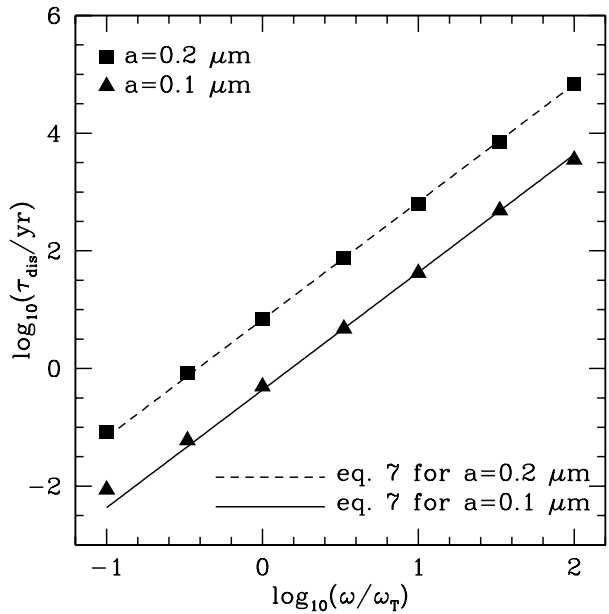}
\caption{
Disalignment time vs. suprathermality for purely insulating grains.
        }
\label{fig:tau_dis_ins}
\end{figure}

Fig. \ref{fig:tau_dis_ins}
displays $\tau_{\mathrm{dis}}$ versus $\omega / \omega_T$ for 
perfectly insulating grains with $a=0.1$ and $0.2 \ \micron$.
For each case, $\tau_{\mathrm{dis}}$ is 
taken to be its average over the 2 realizations.  The solid (dashed)
curves are $\tau_{\mathrm{dis}}$ from equation (\ref{eq:tau_dis1}) for 
$a = 0.1 \ \micron$ ($0.2 \ \micron$).  We employ $\tau_{\mathrm{flip}}$ and 
the average value of $|p_z|$ (for use in evaluating $\Upsilon$) as 
determined from the simulation.  The agreement between the measured 
values of $\tau_{\mathrm{dis}}$ and those calculated 
with equation (\ref{eq:tau_dis1}) is surprisingly good.  The expectation that
$\tau_{\mathrm{dis}} \propto (\omega / \omega_T)^2$ is well confirmed.

The disalignment times found using equation (\ref{eq:tau_dis1}) and 
the values of $\tau_{\mathrm{flip}}$ from the simulations
are substantially shorter than those from W06 (see figs. 2 and 3 in W06); 
the discrepency exceeds 2 orders of magnitude when $a=0.1 \ \micron$.  
Our simulations yield much larger values of $|p_z|_{\mathrm{av}}$ than estimated 
by W06, and $\tau_{\mathrm{dis}} \propto |p_z|_{\mathrm{av}}^{-2}$.  The 
estimate of $\tau_{\mathrm{flip}}$ in W06 is also substantially larger than 
our result.  When equation (\ref{eq:tau_dis1}) is used 
($\tau_{\mathrm{dis}} \propto \tau_{\mathrm{flip}}^{-1}$), this partially 
compensates 
for the difference associated with the $p_z$ estimates.  However, given the
larger estimate for $\tau_{\mathrm{flip}}$, W06 employed equation 
(\ref{eq:tau_dis2}) when $a=0.1 \ \micron$; in this case, $\tau_{\mathrm{dis}}$
is larger by a factor $\approx 2$ when equation (\ref{eq:tau_dis2}) is 
used than when equation (\ref{eq:tau_dis1}) is used.

\begin{figure}
\includegraphics[width=84mm]{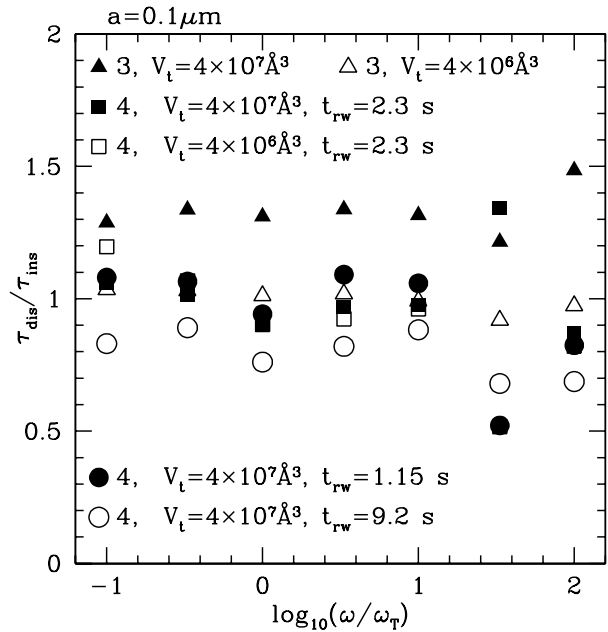}
\caption{
Ratio of disalignment time $\tau_{\mathrm{dis}}$ for various models to that 
for a perfectly insulating grain, $\tau_{\mathrm{ins}}$, for 
$a = 0.1 \ \micron$.  Model
number from \S \ref{sec:idealizations}; volume per deep trap, $V_t$; and 
random walk time, $t_{\mathrm{rw}}$, are indicated.   
        }
\label{fig:ratio_a0.1}
\end{figure}

Fig. \ref{fig:ratio_a0.1} shows the ratio of $\tau_{\mathrm{dis}}$ for several 
simulation runs to its value for the perfectly insulating case, 
$\tau_{\mathrm{ins}}$, for $a = 0.1 \ \micron$.  All of the simulations from 
Table \ref{tab:sim} are included, except for the perfectly conducting case.  
The results always lie within $\approx 50$ percent of unity, with somewhat 
greater scatter when 
$\omega / \omega_T > 10$; the results for these high-$\omega$ cases are not
particularly reliable, since the corresponding $N_{\mathrm{dev}}$ are small
(see Table \ref{tab:sim}).  The ratio 
$\tau_{\mathrm{dis}} / \tau_{\mathrm{ins}}$ also 
lies within 50 percent of unity for the model 3 run for $a = 0.2 \ \micron$.

\begin{figure}
\includegraphics[width=84mm]{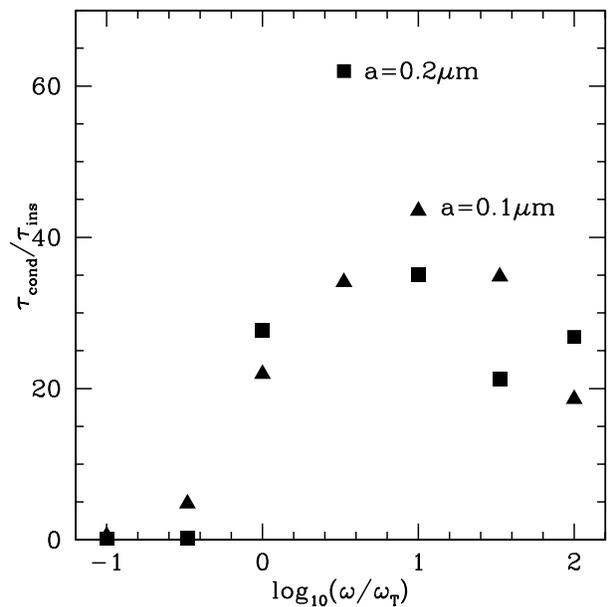}
\caption{
Ratio of disalignment time for a perfectly conducting grain, 
$\tau_{\mathrm{cond}}$,
to that for a perfectly insulating grain, $\tau_{\mathrm{ins}}$, for 
$a = 0.1$ and $0.2 \ \micron$, as indicated.
        }
\label{fig:ratio_cond}
\end{figure}

Fig. \ref{fig:ratio_cond} displays the ratio of $\tau_{\mathrm{dis}}$ for a 
perfectly conducting grain, $\tau_{\mathrm{cond}}$, to $\tau_{\mathrm{ins}}$, 
for $a = 0.1$ and $0.2 \ \micron$.  We assumed that $p_z = 0.1 \, Zea$, which
seems conservative for grains sufficiently asymmetric to produce the observed
polarization.  However, a solution of the electrostatic boundary value 
problem for model aspherical grains would be needed to confirm this choice.
The disalignment times tend to be 1 to 2 orders of magnitude longer for
conducting grains than for insulating grains.  This is not surprising, since
$Z$, and hence $p_z$ for conductors, does not change sign (although it does
fluctuate).  Note that the data points for $\omega / \omega_T \ga 10$ are not 
reliable, given the small values of $N_{\mathrm{dev}}$ in these cases
(Table \ref{tab:sim}).

\section{CONCLUSIONS}
\label{sec:conclusion}

We have conducted a more detailed analysis of grain disalignment 
associated with the time-varying electric dipole moment than was attempted 
in W06, focusing on suprathermally rotating silicate grains.    
We considered 4 idealized models for how charge is transported within the 
grain (\S \ref{sec:idealizations}):  a perfect insulator, 2 models involving
special sites in the grain (`deep traps') where electrons or holes are
effectively trapped, and a perfect conductor.  The resulting disalignment
times $\tau_{\mathrm{dis}}$ for the first 3 models are highly consistent 
(Fig. \ref{fig:ratio_a0.1}) and substantially shorter (up to 2 orders
of magnitude) than those obtained by W06 (cf. Fig. \ref{fig:tau_dis_ins} 
here with figs. 2 and 3 in W06).  We expect the behavior of real grains to
be bracketed by these 3 models.  Disalignment proceeds more slowly (up 
to 2 orders of magnitude; Fig. \ref{fig:ratio_cond}) for conducting grains, 
but we do not expect this idealization to be realistic for interstellar
grains.  

In treating the collisional charging, we neglected the gas-grain drift.
Drift can, in principle, affect the time variation of the electric dipole
moment.  For a non-rotating grain, there may be a stable contribution to
$\bmath{p}$ directed along the drift velocity.  For a grain rotating 
uniformly about $\bmath{\hat{a}_1}$, the charging rate may have some 
dependence on latitude on the grain, suppressing flips in $p_z$.  We examine 
this possibility in Appendix \ref{app:drift} and conclude that flipping is 
not suppressed.  

In the radiative torque alignment scenario, suprathermal rotation with 
$\omega / \omega_T \approx 100$ and alignment times $\ga 10^5 \ \mathrm{yr}$ 
appear to be typical (Draine \& Weingartner 1997; Lazarian \& Hoang 2007;
Hoang \& Lazarian 2008), though additional studies are needed to confirm
these results.  We have found disalignment times $\la 10^5 \ \mathrm{yr}$ 
when $\omega / \omega_T \approx 100$ (Figs. \ref{fig:tau_dis_ins} and 
\ref{fig:ratio_a0.1}), presenting a severe challenge to the radiative torque
model.  

Much of the physics involved in the disalignment mechanism has not been 
directly verified, including the details of the charging and the 
turbulence-induced grain acceleration (Yan et al. 2004).  Perhaps current
models of these processes are incomplete in such a way as to overestimate the
magnitude of the disalignment.  

Alternatively, interstellar grains might
contain superparamagnetic inclusions (Jones \& Spitzer 1967), which could
increase the magnetic susceptibility by orders of magnitude.  The 
parameter $\Upsilon$ would be decreased by the same factor, and the 
disalignment time $\propto \Upsilon^{-2}$ when $\Upsilon \ll 1$ 
(eqs. \ref{eq:tau_dis1} and \ref{eq:tau_dis2}).  Recently, Lazarian \& 
Hoang (2008) found that the presence of superparamagnetic inclusions can
modify alignment by radiative torques, yielding a higher degree of 
alignment than experienced by grains free of inclusions.  Perhaps
superparamagnetic inclusions also suppress drift-induced disalignment.

\section*{ACKNOWLEDGMENTS}

We are grateful to Yuri Mishin for valuable discussions and an anonymous
referee for helpful comments.  
JCW is a Cottrell Scholar of Research Corporation.

\appendix

\section{COLLISIONAL CHARGING FOR A DRIFTING GRAIN}
\label{app:drift}

Consider a grain drifting with velocity $\bmath{v_{\rm gr}}$ with 
respect to the gas.  The grain rotates uniformly about $\bmath{\hat{a}_1}$, 
which is inclined at angle $\theta_{Jv}$ relative to $\bmath{v_{\rm gr}}$.  

To treat the collisional charging in this case,
we first construct a large sphere with radius $r_{\mathrm{big}}$ 
instantaneously centered on the grain.  
Adopting the rest frame of the gas and taking the direction of the drift 
velocity $\bmath{{v}_{\mathrm{gr}}}$ as the polar axis for spherical 
coordinates, the velocity $\bmath{v}$ of a gas-phase particle has components
$(v, \theta_{\mathrm{in}}, \phi_{\mathrm{in}})$.  The rate at which gas-phase 
particles enter the large sphere from within solid angle 
$\mathrm{d}\cos \theta_{\mathrm{in}} \, \mathrm{d} \phi_{\mathrm{in}}$ about 
$(\theta_{\mathrm{in}}, \phi_{\mathrm{in}})$  and with speeds
between $v$ and $v + dv$ is 
\be
dR = \pi r_{\mathrm{big}}^2 n \, \frac{1}{4\pi} \, \mathrm{d}\cos 
\theta_{\mathrm{in}} \, \mathrm{d} \phi_{\mathrm{in}} \, 
P(v) \mathrm{d} v \, |\bmath{v} - \bmath{v}_{\mathrm{gr}} | 
\ee
where $P(v)$ is the Maxwell speed distribution.  After integrating over
$\phi_{\mathrm{in}}$, 
\be
\label{eq:dR_big}
\mathrm{d} R = \pi r_{\mathrm{big}}^2 n \left( \frac{8 k_{\mathrm{B}} T}{\pi m} 
\right)^{1/2} \, \mathrm{d} \tilde{R}
\ee
with 
\be
\mathrm{d} \tilde{R} = \mathrm{d} u \, \mathrm{d} \cos \theta_{\mathrm{in}} \, 
u^2 \, u_1 \, \exp(-u^2) ;
\ee
the dimensionless speed $u = v / v_{\mathrm{th}}$ and 
\be
u_1 = (u^2 + u_{\mathrm{gr}}^2 + 2 u u_{\mathrm{gr}} 
\cos \theta_{\mathrm{in}})^{1/2}
\ee
is the particle's dimensionless speed in the rest frame of the grain 
($u_{\mathrm{gr}} = v_{\mathrm{gr}} / v_{\mathrm{th}}$).  
Integrating over the entire large sphere yields
\be
\label{eq:R_tilde_big}
\tilde{R} = 1 + \frac{u_{\mathrm{gr}}^2}{3} - \frac{u_{\mathrm{gr}}^4}{3} 
\int_0^1 \mathrm{d} x \exp(- u_{\mathrm{gr}}^2 x) (1 - \sqrt{x})^3.
\ee
The arrival angle $\theta_1$ in the rest frame of the grain is characterized
by
\be
\cos \theta_1 = \frac{u \cos \theta_{\mathrm{in}} + u_{\mathrm{gr}}}{u_1}.
\ee

When $v_{\perp} = 1 \ \mathrm{km} \ \mathrm{s}^{-1}$ and 
$T_{\mathrm{gas}} = 100 \ \mathrm{K}$, $u_{\mathrm{gr}} = 0.0182$
for electrons and $0.778$ for protons.  For each of these values
of $u_{\mathrm{gr}}$, we calculate $u_1$ and $\mathrm{d} \tilde{R}$ 
for $10^{10}$ values of $(u, \theta_{\mathrm{in}})$ ($10^5$ for each input 
variable) spaced evenly in 
probability (as described in the text surrounding eq. \ref{eq:u_spacing}) and
with a maximum value of $u = 4$.  The results are used to construct the 
probability $P(u_1)$ that an incoming particle has dimensionless
speed in the grain's frame $\le u_1$ (with 100 bins in $u_1$).  For each
value of $u_1$, the corresponding cumulative probability $P(\cos \theta_1)$
is constructed, again with 100 bins.  Note that the minimum possible value
of $\cos \theta_1$ is $-1$ when $u > u_{\mathrm{gr}}$ and 
$[1-(u/u_{\mathrm{gr}})^2]^{1/2}$ when $u < u_{\mathrm{gr}}$.

To simulate the collisional charging, we adopt a time step 10 times smaller 
than
the inverse of the rate at which electrons enter the large sphere surrounding
the grain (eqs.~\ref{eq:dR_big} and \ref{eq:R_tilde_big}).  
In each step, we draw a random number to determine whether or
not an electron enters the large sphere; likewise for a proton.
When a charged particle enters
the large sphere, a value of $u_1$ is picked randomly from its distribution.
Then, $\cos \theta_1$ is chosen randomly from the distribution for the given
$u_1$.  The final component of the particle's velocity in the grain's rest
frame, $\phi_1$, is selected randomly from a uniform distribution between 0
and $2 \pi$.  Two components of the incoming particle's position remain 
to be determined (given $r = r_{\mathrm{big}}$):  the impact parameter $b$ 
and the
azimuthal angle $\alpha_1$.  These are both chosen randomly ($b$ from a 
uniform distribution in $b^2$).  

At this point, the position and velocity of the incoming particle are specified
relative to a coordinate system at rest with respect to the grain and 
with $\bmath{v}_{\mathrm{gr}}$ as the polar axis 
(`$\bmath{v}_{\mathrm{gr}}$-coordinates').
Denoting Cartesian axes in this coordinate system as $\bmath{\hat{x}_v}$, 
$\bmath{\hat{y}_v}$, and $\bmath{\hat{z}_v}$, the velocity is given by
\be
\label{eq:v1}
\bmath{v_1} = -v_{\mathrm{th}} u_1 (\bmath{\hat{x}_v} \sin \theta_1 \cos \phi_1 
+ \bmath{\hat{y}_v} \sin \theta_1 \sin \phi_1 + \bmath{\hat{z}_v}  
\cos \theta_1)
\ee
and the position by
\begin{eqnarray}
\nonumber
x_v & = & b \cos \alpha_1 \cos \theta_1 \cos \phi_1 - 
b \sin \alpha_1 \sin \phi_1\\
& & + z_{\mathrm{arr}} \sin \theta_1 \cos \phi_1
\end{eqnarray}
\begin{eqnarray}
\nonumber
y_v & = & b \cos \alpha_1 \cos \theta_1 \sin \phi_1 + 
b \sin \alpha_1 \cos \phi_1\\
& & + z_{\mathrm{arr}} \sin \theta_1 \sin \phi_1
\end{eqnarray}
\be
\label{eq:z_v}
z_v = - b \cos \alpha_1 \sin \theta_1 + z_{\mathrm{arr}} \cos \theta_1
\ee
where $z_{\mathrm{arr}} = (r_{\mathrm{big}}^2 - b^2)^{1/2}$.  

Denoting a Cartesian coordinate system attached to the grain body by 
$(x_J, y_J, z_J)$, 
\be
\label{eq:hat_z_J}
\bmath{\hat{z}_J} = \bmath{\hat{a}_1} = \bmath{\hat{z}_v} \cos \theta_{Jv} + 
\bmath{\hat{x}_v} \sin \theta_{Jv} 
\ee
\be
\bmath{\hat{x}_J} = (\bmath{\hat{x}_v} \cos \theta_{Jv} - \bmath{\hat{z}_v} 
\sin \theta_{Jv}) \cos \lambda + \bmath{\hat{y}_v} \sin \lambda
\ee
\be
\bmath{\hat{y}_J} = \bmath{\hat{y}_v} \cos \lambda - (\bmath{\hat{x}_v} 
\cos \theta_{Jv} - \bmath{\hat{z}_v} \sin \theta_{Jv}) \sin \lambda
\ee
where $\lambda$ is the phase angle of the grain's rotation and is selected
randomly.  (We neglect the rotation of the grain during the approach of the
gas-phase particle.)  Cartesian axes with $\bmath{p}$ as the polar axis 
(`$\bmath{p}$-coordinates') are given by
\be
\bmath{\hat{z}_p} = \bmath{\hat{p}} = \bmath{\hat{x}_J} \sin \theta_{pJ} 
\cos \phi_{pJ}  + \bmath{\hat{y}_J} \sin \theta_{pJ} \sin \phi_{pJ} + 
\bmath{\hat{z}_J} \cos \theta_{pJ} 
\ee
\be
\bmath{\hat{x}_P} = \bmath{\hat{x}_J} \cos \theta_{pJ} \cos \phi_{pJ} + 
\bmath{\hat{y}_J} \cos \theta_{pJ} \sin \phi_{pJ} - 
\bmath{\hat{z}_J} \sin \theta_{pJ}
\ee
\be
\label{eq:hat_y_p}
\bmath{\hat{y}_P} = - \bmath{\hat{x}_J} \sin \phi_{pJ} + 
\bmath{\hat{y}_J} \cos \phi_{pJ}.
\ee
From equations (\ref{eq:hat_z_J})--(\ref{eq:hat_y_p}), 
we find the following dot products for use in transforming the position and
velocity of the incoming gas-phase particle from 
$\bmath{{v}_{\rm gr}}$-coordinates (eqs. \ref{eq:v1}--\ref{eq:z_v}) to 
$\bmath{p}$-coordinates: 
\be
\bmath{\hat{x}_p \cdot \hat{x}_v} = 
\cos \theta_{pJ} \cos \theta_{Jv} \cos (\phi_{pJ} 
+ \lambda) - \sin \theta_{pJ} \sin \theta_{Jv}
\ee
\be
\bmath{\hat{x}_p \cdot \hat{y}_v} = 
\cos \theta_{pJ} \sin (\phi_{pJ} + \lambda)
\ee
\be
\bmath{\hat{x}_p \cdot \hat{z}_v} = - \cos \theta_{pJ} \sin \theta_{Jv} 
\cos (\phi_{pJ} + \lambda) - \sin \theta_{pJ} \cos \theta_{Jv}
\ee
\be
\bmath{\hat{y}_p \cdot \hat{x}_v} = 
- \cos \theta_{Jv} \sin (\phi_{pJ} + \lambda)
\ee
\be
\bmath{\hat{y}_p \cdot \hat{y}_v} = \cos (\phi_{pJ} + \lambda)
\ee
\be
\bmath{\hat{y}_p \cdot \hat{z}_v} = \sin \theta_{Jv} \sin (\phi_{pJ} + \lambda)
\ee
\be
\bmath{\hat{z}_p \cdot \hat{x}_v} = \sin \theta_{pJ} \cos \theta_{Jv} 
\cos (\phi_{pJ} + \lambda) + \cos \theta_{pJ} \sin \theta_{Jv}
\ee
\be
\bmath{\hat{z}_p \cdot \hat{y}_v} = \sin \theta_{pJ} \sin (\phi_{pJ} + \lambda)
\ee
\be
\bmath{\hat{z}_p \cdot \hat{z}_v} = - \sin \theta_{pJ} \sin \theta_{Jv} 
\cos (\phi_{pJ} + \lambda) + \cos \theta_{pJ} \cos \theta_{Jv}.
\ee

Finally, the Cartesian $\bmath{p}$-coordinates of the incoming particle's 
position and velocity are used in the following geometric relations to find 
the components in spherical coordinates:
\be
r = r_{\mathrm{big}}
\ee  
\be
\theta = \cos^{-1} (z_p / r)
\ee
\be
\phi = 2 \tan^{-1} \left( \frac{r_{p0} - x_p}{y_p} \right)
\ee
\be
\frac{\mathrm{d} r}{\mathrm{d} t} = \frac{1}{r} \left( x_p 
\frac{\mathrm{d} x_p}{\mathrm{d} t} + y_p \frac{\mathrm{d} y_p}{\mathrm{d} t} 
+ z_p \frac{\mathrm{d} z_p}{\mathrm{d} t} \right)
\ee
\be
\frac{\mathrm{d} \theta}{\mathrm{d} t} = - \frac{1}{r r_{p0}} \left(
r \frac{\mathrm{d} z_p}{\mathrm{d} t} - z_p \frac{\mathrm{d} r}
{\mathrm{d} t} \right)
\ee
\be
\frac{\mathrm{d} \phi}{\mathrm{d} t} = \frac{1}{r_{p0}^2} \left( x_p 
\frac{\mathrm{d} y_p}{\mathrm{d} t} - y_p \frac{\mathrm{d} x_p}
{\mathrm{d} t} \right)
\ee
with $r_{p0} = (x_p^2 + y_p^2)^{1/2}$.
The critical impact parameter $b_{\mathrm{crit}}$ depends on the particle speed
$v_1 = v_{\mathrm{th}} u_1$ and 
$\cos \theta_0 = - v_1^{-1} \mathrm{d}z_p/\mathrm{d}t$ 
(eq. \ref{eq:b_crit}).  If $b \le b_{\mathrm{crit}}$, then we integrate the 
equations of motion (\ref{eq:r_dot_dot})--(\ref{eq:phi_dot_dot}) to 
determine where on the grain surface the particle hits.  

We have tried various values of $r_{\mathrm{big}}$.  Of course, larger values 
yield higher accuracy but also require smaller time steps.  We found that 
$r_{\mathrm{big}} = 50 a$ yields high accuracy and is not prohibitively time 
consuming.

Substituting the collisional charging procedure described here in our 
charging simulations (and including both electrons and protons), we examined 
a perfectly insulating grain with 
$a = 0.1 \ \micron$.  With a duration of $100 \ \mathrm{yr}$, we found that 
$|p_z|_{\mathrm{av}}/ea \approx 2.5$ and $\tau_{\mathrm{flip}}$ ranges from 
$5.8 \times 10^{-4}$ to $6.0 \times 10^{-4}$ as $\cos \theta_{Jv}$ ranges
from 0 to 1.  These are very close to the results obtained previously, 
neglecting grain drift (and ignoring protons) in the treatment of collisional 
charging (see Table \ref{tab:sim}).  
Due to precession of $\bmath{J}$ about $\bmath{B}$, the angle 
$\theta_{Jv}$ changes on a time-scale short compared with the simulation time 
of $100 \ \mathrm{yr}$ (but an order of magnitude longer than 
$\tau_{\mathrm{flip}}$).   
However, our results imply that the behavior of the electric dipole moment is 
insensitive to the value of $\theta_{Jv}$.  Thus, we conclude that the 
neglect of grain drift in \S \ref{sec:results} does not yield significant
error.

\end{document}